\documentclass[conference]{IEEEtran}
\IEEEoverridecommandlockouts
\usepackage{cite}
\usepackage{amsmath,amssymb,amsfonts}
\usepackage{algorithmic}
\usepackage{textcomp}
\usepackage[keeplastbox]{flushend}
\usepackage{caption,subcaption}
\usepackage{graphicx}
\usepackage{multirow}
\begin{document}

\title{IoT Security*\\
{\footnotesize \textsuperscript{*}Note: Sub-titles are not captured in Xplore and
should not be used}
\thanks{Identify applicable funding agency here. If none, delete this.}
}

\title{Device Authentication Codes based on RF Fingerprinting using Deep Learning}


\author{\IEEEauthorblockN{Joshua~Bassey,~Xiangfang Li, Lijun Qian}
\IEEEauthorblockA{CREDIT Center and Department of Electrical and Computer Engineering \\
Prairie View A\&M University, Texas A\&M University System  \\
Prairie View, TX 77446, USA \\
Email: jbassey, xili, liqian@pvamu.edu}
}


\maketitle

\begin{abstract}
In this paper, we propose Device Authentication Code (DAC), a novel method for authenticating IoT devices with wireless interface by exploiting their radio frequency (RF) signatures. The proposed DAC is based on RF fingerprinting, information theoretic method, feature learning, and discriminatory power of deep learning. Specifically, an autoencoder is used to automatically extract features from the RF traces, and the  reconstruction error is used as the DAC and this DAC is unique to the device and the particular message of interest. Then Kolmogorov-Smirnov (K-S) test is used to match the distribution of the reconstruction error generated by the autoencoder and the received message, and the result will determine whether the device of interest belongs to an authorized user. We validate this concept on two experimentally collected RF traces from six ZigBee and five universal software defined radio peripheral (USRP) devices, respectively. The traces span a range of Signal-to-Noise Ratio by varying locations and mobility of the devices and channel interference and noise to ensure robustness of the model. 
Experimental results demonstrate that DAC is able to prevent device impersonation by extracting salient features that are unique to any wireless device of interest and can be used to identify RF devices. Furthermore, the proposed method does not need the RF traces of the intruder during model training yet be able to identify devices not seen during training, which makes it practical.  
\end{abstract}

\begin{IEEEkeywords}
RF fingerprinting, Deep Learning, Internet of Things, autoencoder, Kolmogorov-Smirnov (K-S) test.
\end{IEEEkeywords}

\section{Introduction}

We introduce a novel method for authenticating IoT devices with wireless interfaces based on their radio frequency (RF) signatures called the Device Authentication Code (DAC). DAC exploits the potential of RF fingerprinting and the feature learning power of autoencoders. DAC similar to the message authentication code (MAC) approach to message authentication in cryptographic network security applications. With the DAC any IoT device with a wireless interface that transmits a wireless can be authenticated and the integrity of the transmitted signal verified because the method exploits the features inherent to the device alone that uniquely distinguishes that device from any other device. 

\begin{figure*}
	 \centering
    	 \includegraphics[width=0.9\textwidth]{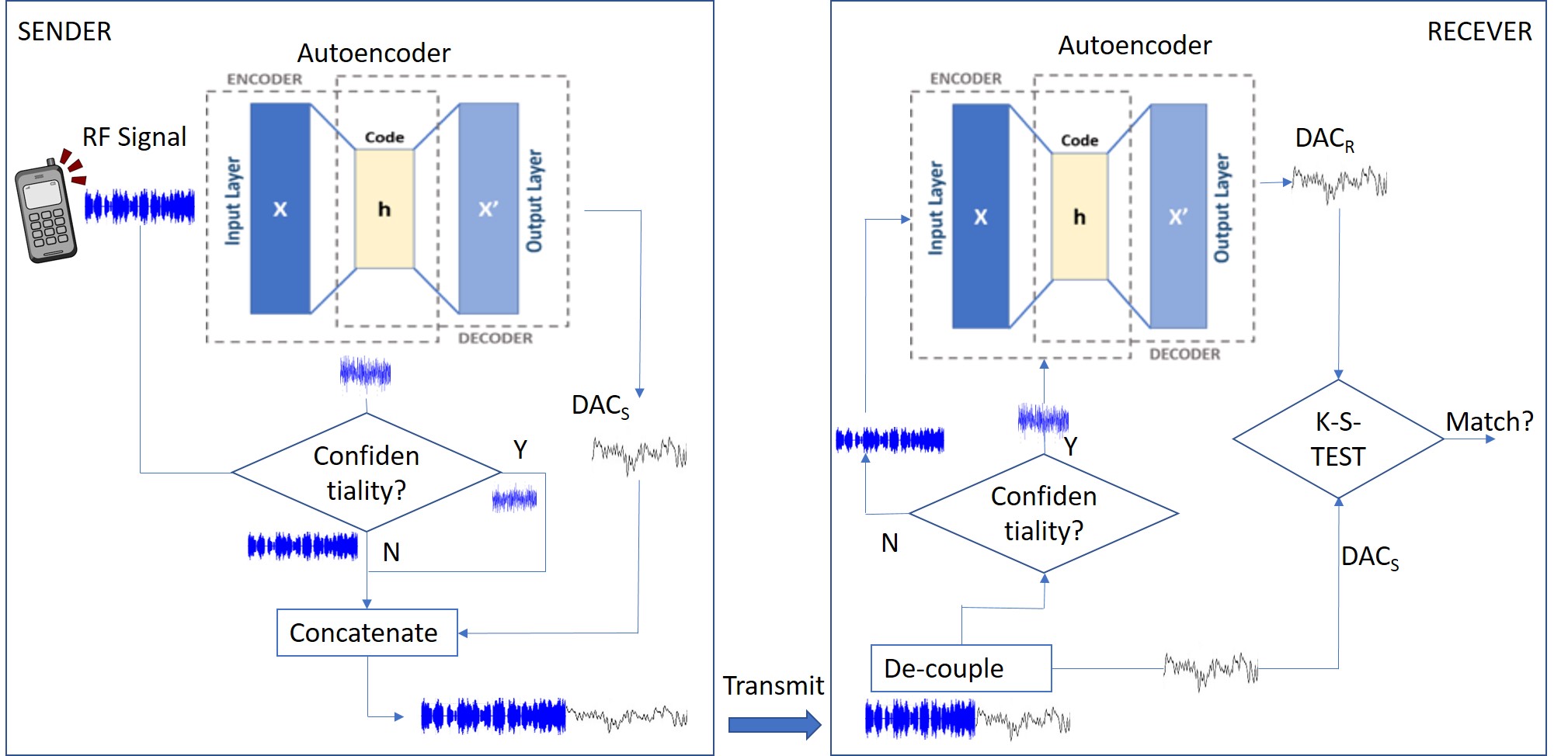}
      	\caption{Authentication and Confidentiality using Device Authentication Code}
    \label{fig:DAC_Model}
\end{figure*}

The process of device authentication using the DAC scheme  is depicted in Figure \ref{fig:DAC_Model}, an AE is trained to reconstruct the inputs given to it by minimizing the reconstruction error. This reconstruction error is used as the device authentication code (DAC). A device passes the signal to be transmitted through a pretrained auto-encoder (AE) based model.  The signal is concatenated with  the reconstruction (DAC) and transmitted. At the receiver, the received signal is decoupled into the original signal and DAC. The signal is then passed through the AE model deployed at the receiver to generate another DAC. Device authentication is done by comparing the two DAC using the Kolmogorov-Smirnov statistic. An exact match in both DAC authenticates the transmitting device. If confidentiality is required, the encoded version of the signal could be transmitted instead of the raw signal.

The identification of  IoT devices based on the physical characteristics of their built-in components is applicable not just for authentication but also for tracking both the device and its user~\cite{paper6_Baldini2017}. The fingerprints are typically created during the manufacturing process as the base materials of the components are created. Creation of fingerprints are usually accidental or inherent to the process. However, it is possible to generate and insert them on purpose. In either case, the RF fingerprints are a result of minute variations in the electronic components~\cite{paper6_Baldini2017}. When appropriately analyzed, RF fingerprints can be exploited to identify and distinguish one device from another, even from the same manufacturer model~\cite{paper6_Baldini2017}. Unlike RF features, identifiers at other layers such as MAC addresses and International Mobile Subscriber Identity (IMSI) are relatively easy to impersonate~\cite{Tomko2006, AnIntroducitiontoDLShea}.

In our previous work~\cite{Bassey2019}, we proposed an intrusion detection model pipeline based on RF fingerprinting combining deep learning, dimension reduction and clustering models. After training, the model pipeline clusters the RF traces from authorized devices into separate clusters, corresponding to each authorized device. When RF traces from an unauthorized device goes through the pipeline, it is clustered as a new cluster (the intruder). This framework can the detect presence of an intruder by determining the amount of unique clusters (devices) in the perimeter of interest. However, this model cannot ascertain which RF trace belongs to an intruder. The DAC is a more holistic and robust approach that mitigates this limitation.


The rest of the paper is structured in the following manner: In Section \ref{sec:background}, we present  background  on components from which the inspiration of our work is drawn. An explanation of the proposed approach  is given in Section \ref{sec:poposedSolution}.  In Section \ref{sec:experiments} information on experiments including data collection and analysis of results are presented. Related work are presented in Section \ref{sec:discussion} and Section \ref{sec:conclusions} contains conclusions and future works.

\section{Background}
\label{sec:background}

Device identification via RF fingerprinting typically involves data collection, processing, feature extraction and device identification. In this work we apply deep learning to automate these tasks. This is because its success in feature learning, and classification accuracy across multiple domains such as computer vision, speech, natural language and signal processing. However, in supervised deep learning samples and labels from all classes of interest must be present during training. During inference, a test sample from a class not observed during training will be classified as one of the already seen classes. For this reason we adopt unsupervised learning.


Message authentication comprises methods used to verify the identity of the sender and/or the integrity of the message (meaning it has not been modified, deleted or is being replayed). Traditionally, the main methods for performing authentication are: message encryption, the use of message authentication codes (MAC) and hash functions. In this section we briefly introduce the MAC approach.

\subsection{Message Authentication Code (MAC)}
\label{subsec: MAC}
Consider sending a message $M$ between two parties \textit{A} and \textit{B} that share a secret key \textit{K}. To transmit a message to \textit{B}, \textit{A} uses \textit{K} to create a message authentication code ($MAC_{S}$), a fixed sized cryptographic checksum and function $MAC = C(K,M)$ of the message and the shared key. The MAC is appended to the message and transmitted. \textit{B} applies the MAC function on the message and generates a new $MAC_{R}$ using the secret key. The newly generated $MAC_{R}$ is compared with the received $MAC_{S}$. If $MAC_{S} = MAC_{R}$, then:

\begin{enumerate}
\item \textit{B} is assured that $M$ has not been altered because if an intruder modifies $M$ without modifying $MAC_{S}$, then $MAC_{R}$ will not match $MAC_{S}$. Also, the intruder cannot modify $MAC_{S}$ to reflect changes in the message because he does not have $K$.
\item \textit{B} is also assured that the message came from \textit{A} because only \textit{A} has \textit{K} required to generate a message with the correct MAC.
\end{enumerate}

The above scheme describes the approach to authentication. There is no confidentiality because anyone can have access to $M$. Traditionally, confidentiality is introduced by encrypting $M$ either before or after the MAC procedure. The later is the popular choice. However both parties need two set of keys.

\subsection{Autoencoders}

\begin{figure}
	 \centering
    	 \includegraphics[width=0.4\textwidth]{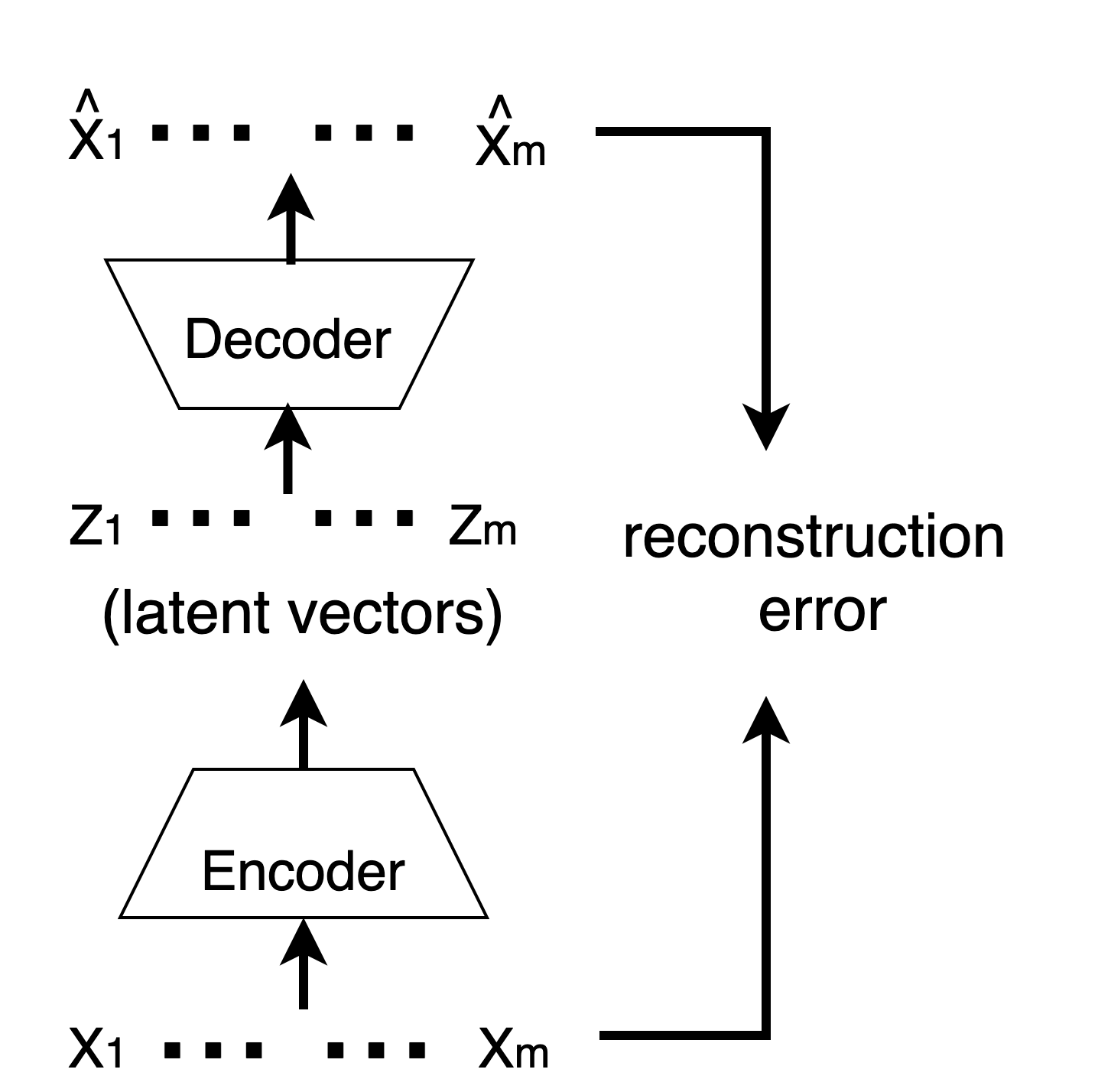}
      	\caption{Architecture of an Autoencoder}
    \label{fig:AE_arch}
\end{figure}

Autoencoders (AE) are neural networks with the objective of reconstructing data input into them (Figure \ref{fig:AE_arch}). Mathematically, the autoencoder attempts to learn the identity function:
\begin{equation}
f_{W,b}(x) = x
\end{equation}
by minimizing the ``reconstruction error" between the input and its reconstruction given as:
\begin{equation}
L(x,x^{'}) = \vert \vert x - x^{'} \vert \vert^{2}
\end{equation}

First the AE learns an ``encoded'' representation of the data, by extracting the inherent structure in the data~\cite{AE_Review_Tschannen_2018}. Learning the encoded representation can be achieved by restricting the number of nodes in the encoding layers as in undercomplete autoencoders~\cite{book1-goodfellow2016}. Overcomplete autoencoders learn structure by imposing  other regularization constraints on the encoding layer such as sparsity as in sparse autoencoders~\cite{paper56_shahin2017}, or addition of noise as in denoising autoencoders~\cite{paper55_Vincent2010}. 

Convolutional autoencoders (CAE)  exploit spatial relationships in data by weight sharing~\cite{paper58_Jonathan2011}. AEs can be extended to make deeper networks and can be trained in a greedy layer-wise manner, where each layer is the latent representation of an already trained AE.

\begin{figure}
	\centering           
	\includegraphics[width=0.5\textwidth]{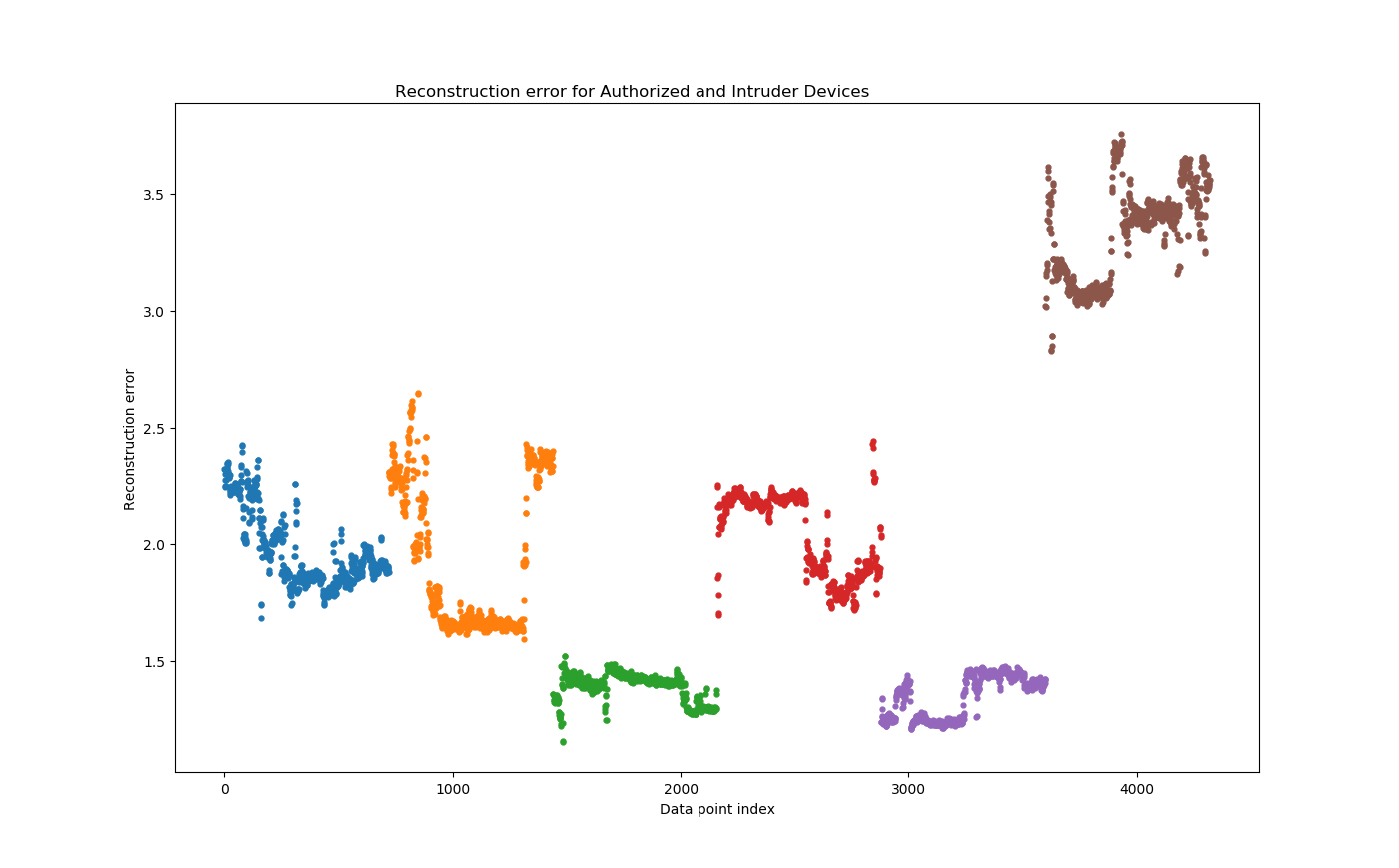}  
	\caption{Reconstruction error for six identical RF devices transmitting identical signals}
	\label{fig:CAE_model}
\end{figure}

\subsection{Kolmogorov-Smirnov (K-S) test}
The K-S test is a non-parametric test used to ascertain whether a sample comes from a population whose distribution is known, or whether the distribution of two populations are the same. In the one-sample test, a one-dimensional probability distribution is compared to a reference probability distribution. In the two-sample test two samples from two distributions are compared. If we define the empirical distribution function (EDF) $F_{n} $ for  \textit{n} independent and identically distributed (i.i.d) observations $X_{i}$ which are ordered as:

\begin{equation}
F_{n}(x) = \frac{1}{n} \sum^{n}_{i=1} I_{[- \infty,x]}(X_{i})
\end{equation}

where $I_{[- \infty,x]}(X_{i})$ is an indicator function equals 1 when $X_{i} \leq x$ and 0 otherwise. Then the K-S statistic for another EDF $F(x)$ is:

\begin{equation}
D_{n} = \sup_{x} \vert F_{n}(x) - F(x) \vert
\end{equation}

where $\sup_{x}$ is the supremum function of the set of distances. The K-S statistic converges to 0 as $\textit{n}$ goes to infinity if the sample is from the distribution $F(x)$. Analogously, for the two-sample test , given two empirical distributions $F_{1,n}$ and $F_{2,m}$ with sample sizes of $n$ and $m$, respectively, the K-S statistic for the first and second sample is 

\begin{equation}
D_{n,m} = \sup_{x} \vert F_{1,n}(x) - F_{2,m}(x) \vert
\end{equation}

Given a specified level $\alpha$, the null hypothesis can be rejected for large sample sizes if

\begin{equation}
D_{n,m} > c(\alpha) \sqrt{\frac{n + m}{nm}}
\end{equation}

where in general 
\begin{equation}
c(\alpha) = \sqrt{- \frac{1}{2} ln \alpha}
\end{equation}

It is possible to set confidence limits on $F(x)$ such that for the test statistic $D_{\alpha}$, if $P(D_{n} > D_{\alpha}) = \alpha$ then $F(x)$ will be contained in $F_{n}(x)$ and a tolerance of width $\pm D_{\alpha}$ with a probability of 1 - $\alpha$. The null hypothesis is that both samples are drawn from the same distribution and the \textit{p}-value is a measure of similarity. If the \textit{p}-value is ``small'', the null hypothesis should be rejected. The KS-test measures the distance between the empirical distribution functions of both samples without any assumptions about the distribution of the data. Unlike the \textit{t}-test, K-S test is robust to scale changes and it is not restricted to identifying changes only in  the mean.


\section{Proposed Solution - Device Authentication Codes (DAC)}
\label{sec:poposedSolution}

\subsection{Problem Statement}

\begin{figure}
	 \centering
    	 \includegraphics[width=0.5\textwidth]{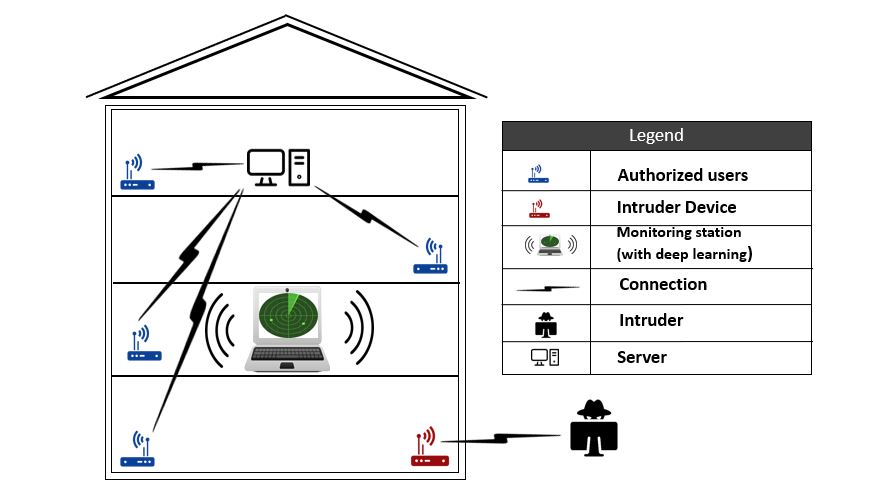}
      	\caption{Problem Scenario}
    \label{fig:DAC_Scenario}
\end{figure}

We herein restate the problem for the purpose of emphasis. There are \textit{n} RF devices with wireless interfaces. All devices are of the same make and model and are considered identical (figure \ref{fig:DAC_Scenario}). All devices are made to transmit the same information. This procedure is repeated but at different noise levels. One of the devices is considered an intruder. Another constraint is that the RF traces from the intruder device are not available for the training of the model. The objective is to authenticate any device of interest and identify the intruder. 

\subsection{Mathematical formulation}
We represent the manufacturing process for a batch of RF devices such as sensors as:
\begin{equation}
S = S_{o} + \mu_{M}
\end{equation}

$S_{o}$ represents the features common to every device in the batch and is required for any device to pass quality control. 
$\mu_{M}$ accounts for minor differences and uncertainties due to the imperfection of the manufacturing process. An autoencoder as a function:

\begin{equation}
\label{eqn:AE_mapping}
X \rightarrow Z \rightarrow \hat{X}
\end{equation}
is a mapping  that encodes an input $X$ to a latent representation $Z$, and decodes $Z$ to recover $X$. Since there is no explicit formula for $\hat{X}$, what has been done instead is to minimize:

\begin{equation}
\label{eqn:AE_MSE}
min \vert \vert X - \hat{X} \vert \vert ^{2}
\end{equation}
during training using data. In wireless communications there are also environmental factors $\varepsilon$, such as channel fading, thermal noise, and effects of device mobility, that are superimposed on the received signal. Therefore the training data can be denoted as:

\begin{equation}
\label{eqn:AE_DATA}
X = f(S, \varepsilon), 
\end{equation}

which is a mapping generated by the underlying stochastic process that consists of $S$ which contains the manufacturing uncertainty $\mu_{M}$, and $\varepsilon$ which contains environmental uncertainty. Based on the premise of RF fingerprinting, $\mu_{M}$ is unique to every RF device. Hence, for a batch of devices:

\begin{equation}
S_{i} = S_{0} + \mu_{M_{i}} \qquad i = 1,...N.
\end{equation}
it is possible to identify $S_{i}$ $\forall i \in N$ using a method that is robust to the environmental uncertainty $\varepsilon$ that affects the data $X$ obtained from the batch of devices. We show experimentally that this can be achieved using an autoencoder and a two-sided K-S test.

\subsection{Intrusion Detection with Autoencoder and K-S test}

Autoencoder based models have been used for anomaly and novelty detection in other domains~\cite{Xia2015},~\cite{AE_Review_Dong_2018}. Autoencoders were used to detect abnormalities in machines by detecting abnormal operation sounds~\cite{Abdet_Dong2018}, and to detect anomalies in video frames~\cite{Abdet2_Gutoski2017}. The idea is based on the fact that a trained autoencoders will produce a low reconstruction error for data from the same or similar distribution as the training data but a high reconstruction error otherwise. Hence the reconstruction error is thresholded and used to identify anomaly.


However, for our specific problem of interest, the devices are of the same make and model, and may transmit identical signals (see Figure~\ref{fig:receivedsignal}). In this case, the threshold approach does not work. Figure \ref{fig:CAE_model} shows the reconstruction error during inference for a CAE trained on five out of six devices with RF traces from one device left out. It is obvious that identifying a novel device (say, any one of the six devices not used in training) with a single threshold will not suffice. Instead we require a metric that can capture and differentiate between distributions.

\begin{figure*}
	 \centering
    	 \includegraphics[width=0.9\textwidth]{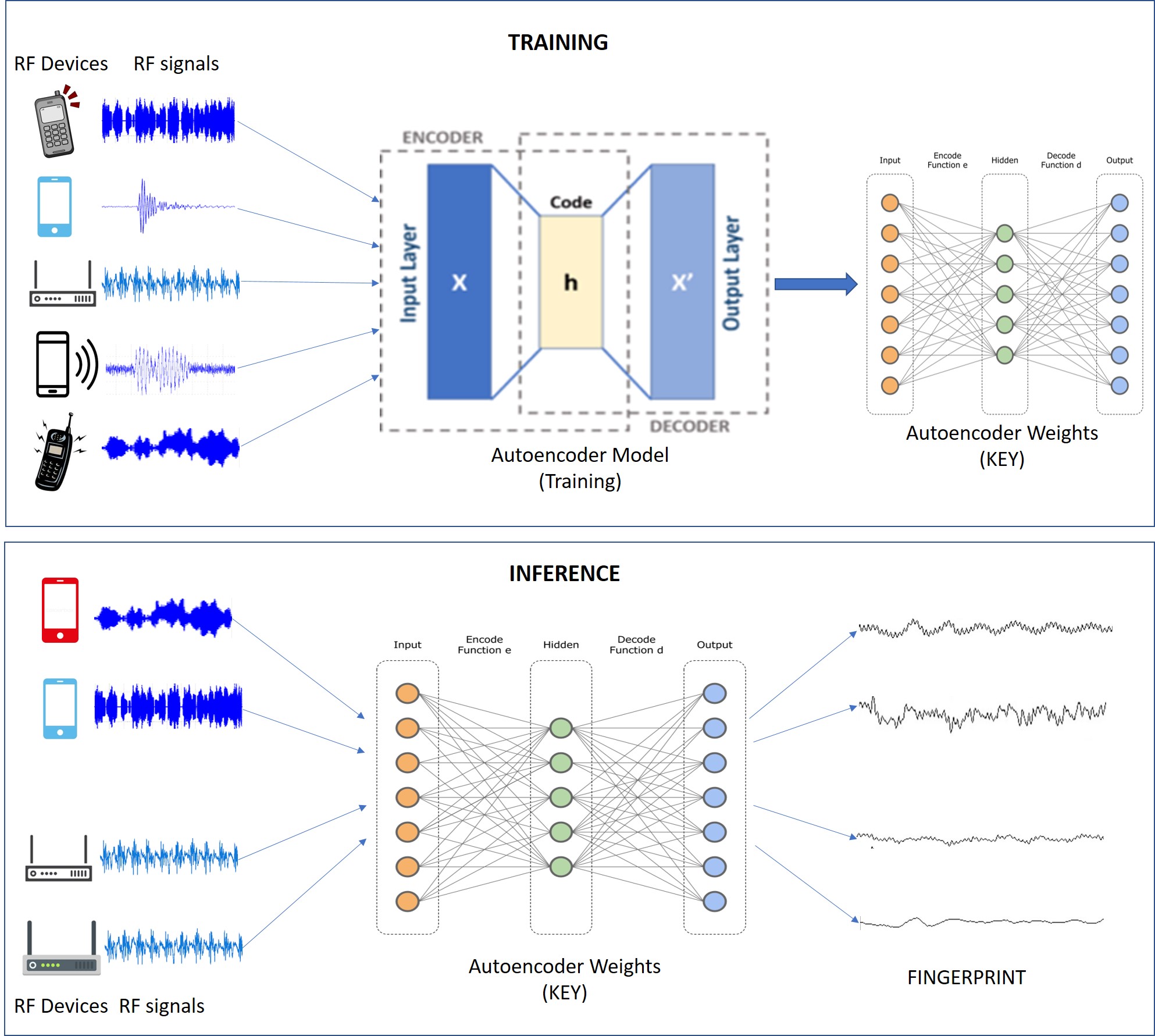}
      	\caption{Training and Inference using DAC}
    \label{fig:DAC_Method}
\end{figure*}

Figure \ref{fig:DAC_Method} depicts the DAC process. An autoencoder is trained on the RF traces from authorized devices. During inference, the distribution of the mean square error (MSE) between signals and their reconstructions will be unique to each device. This holds true for devices of the same make and model, and transmitting identical signals. The MSE are analogous to a fingerprint, and we use them as the DAC in this study. 

For an application such as device registration on a network, the weights of the trained AE is deployed in each authorized device as the secret key. Every device will be required to transmit a predetermined signal upon start-up. The start-up signal is concatenated with its DAC before transmission. A similar setup applies for message authentication for intrusion detection except that any signal can be transmitted by a registered device. The receiving device decodes the signal and reconstructs the original signal using its own key. The DAC is compared to the DAC received from the sending device. A match (i.e., a K-S statistic of 0 and \textit{p}-value of 1) means that the device is an authorized (pre-registered) device, otherwise the device is flagged as a new device (possibly an intruder). 

In the device registration scenario, even if an intruder knows the signal being transmitted and attempt to use a device of the same make and model to transmit an identical signal, his DAC will not be a match at the receiver because the intruder does not have the exact same AE weighs (key). Furthermore, if the message to be transmitted requires confidentiality, the encoded form of the message obtained from the encoder of the autoencoder can be transmitted instead of the original signal. This adds confidentiality and another layer of security. This approach is very secure because the probability of obtaining the exact set of network weights as installed in the device is very low.

\section{Experimental Results}
\label{sec:experiments}

\subsection{Experimental setup}
\label{subsecsec:setup}

\subsubsection{Data Collection}
\label{sec:dataset}

RF traces were collected from two types of devices: (1) six MICAz-MPR2400 sensors, with IEEE 802.15.4 compliant ZigBee ready transceiver operating in [2.4, 2.8] GHz range. (2) NI USRP-293x Software Defined Radio (SDR) Devices. An NI USRP-293x configured in receiver mode using LabVIEW was used to capture the RF traces. All ZigBee devices were configured to transmit at 0, -1, -5, -10, -15dBm SNR, and the USRPs to transmit at [-10dB, 10dB] in steps of 2dB. This is done to simulate a real life scenario which encompasses the effect of degradation and distance, varying channel conditions, noise and device mobility on the signal strength. Transmission were made with the devices mobile as well as stationary. 

\begin{figure}
    	 \includegraphics[width=0.5\textwidth]{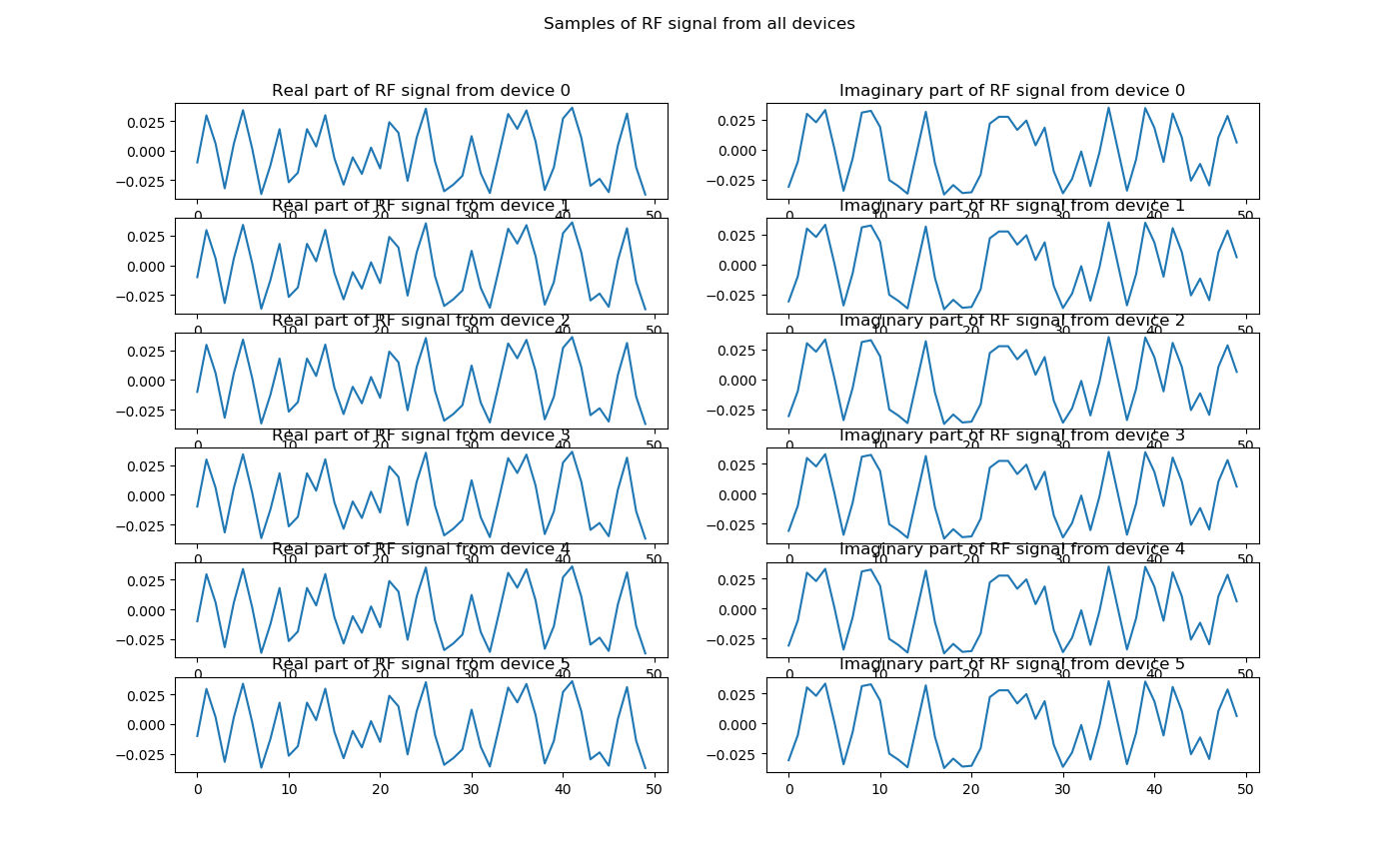}
      	\caption{Sample of RF I and Q data captured from all ZigBee devices }
    \label{fig:receivedsignal}
\end{figure}

RF traces from one of the devices are not used for training the model (figure \ref{fig:DAC_Scenario}). The RF traces consist In-phase (I) and Quadrature (Q) vectors. A sample of the identical RF traces obtained from the ZigBee devices is shown in figure~\ref{fig:receivedsignal}.

\subsubsection{Training and Authentication}
\label{sec:training}

The raw RF traces from the authorized devices is used to train the AE-based model. The model performs automated feature extraction and dimension reduction on the data and during inference, a K-S test is performed on any RF trace to ascertain if it is from an authorized device.

The performance of the proposed approach is evaluated on the generated datasets. As previously stated, one device is considered the intruder and left out of training. $90\%$ of the IQ samples from the other authorized devices at varying SNR levels are used for training, half of the remaining $10\%$ are used for validation and the remaining $5\%$ of samples are mixed with IQ samples from the intruder class for testing. It is worth mentioning again that the data are collected for each device at different SNR levels and combined in order to mimic multipath effects, variation in channel conditions as well as noise. 

\subsection{Results and Analysis}
\label{sec:dataset}
Table \ref{table:raw_data_KS_Statistic} shows the KS statistic and p-values when comparing the DAC of the raw RF traces of all devices. We can consider the rows and columns to represent the sending and receiving devices respectively. For example in row one, the DAC for RF traces from device 0 is compared with DAC computed at the receiver on the RF traces obtained from every device including device 0 itself. The same is done every RF device of interest. The first and second element of the tuple in every cell of the table is the K-S statistic and \textit{p}-value of the K-S test respectively. As previously mentioned, the null hypothesis is that both samples are from the same distribution. 

The null hypothesis cannot be rejected if either the statistic is very low, or the p-value is high. Typical values considered as low are in the range $ [0, 0.1]$. On the other hand, the null hypothesis should be rejected if the K-S statistic is high and the p-value is very low. In other words, a K-S statistic in the range [0, 0.1] and \textit{p}-value in the range [0.9, 1.0] indicates that the device of interest is an authorized device. On the other hand, for a K-S statistic of greater than 0.1 and \textit{p}-value less than 0.9 indicate  that the DAC is from a different distribution and the device of interest is not authorized.

The values of interest in Table  \ref{table:raw_data_KS_Statistic} are highlighted in bold. The first thing to observe is that every cell in the diagonal of the table contains $(0.00, 1.00)$. This indicates a perfect match and it is intuitive since the RF trace is from the authorized device.  Secondly for every other cell, the first element of every tuple are very small values. This shows that the raw RF traces from all the devices according to the K-S test are considered almost identical. Hence the performance of the DAC approach will be based on how far away from zero the K-S statistic is or how close to one the \textit{p}-value is. Only an authorized device will have the values $(0.00, 1.00)$ signifying a DAC match.

Table \ref{table:CAE1024_data_KS_Statistic} show a similar comparison done in Table \ref{table:raw_data_KS_Statistic} but on the DAC obtained for every device using our approach. It can be observed that the values for the K-S statistic here is much higher than in table \ref{table:AE1024_data_KS_Statistic} for the raw RF traces. In fact only one comparison has a K-S statistic of approximately 0.2. If we consider 0.1 to be the minimum threshold below which we cannot reject the null hypothesis, then the CAE model performs quite well and is able to produce very discriminatory features, even though the original data (RF traces) are identical. 

Tables \ref{table:NoiseCAE1024_data_KS_Statistic} and \ref{table:NoiseCAE1024_USRPdata_KS_Statistic} show the performance of the CAE model trained on ZigBee and USRP RF traces respectively collected at single noise levels as well as on RF traces from all noise levels mixed together. The device of interest is device 5 and 2 for the ZigBee and USRP devices respectively. As expected, the model performs better when tested at single noise levels especially for the ZigBee devices. However the model still performs very well on data containing RF traces from all noise levels. This is important because in real life scenario, the wireless signals will seldom be at one noise level due to environmental factors. Hence it is important that the model is robust to these different phenomena.

\begin{table*}
\centering
\caption{Statistic and P-value for raw data from RF Devices.}
\label{table:raw_data_KS_Statistic}
\begin{tabular}{|c|c|c|c|c|c|c|}
\hline
	 		& \multicolumn{6}{c|}{RF Device} 	 \\ \cline{2-7}

{Device of interest}		&  Device 0	& 	Device 1	&  Device 2	&  Device 3	&  Device 4	&  Device 5  \\\hline
 Device 0 		&    (\textbf{0.00, 1.00}) 	& (\textbf{0.02}, 0.00) & (\textbf{0.05}, 0.00) & (\textbf{0.03}, 0.00) & (\textbf{0.04}, 0.00)& (\textbf{0.05}, 0.00)		\\ \cline{1-7}
                			Device 1	&  	 (\textbf{0.02}, 0.00) 	& (\textbf{0.00, 1.00})& (\textbf{0.06}, 0.00) & (\textbf{0.03}, 0.00) & (\textbf{0.05}, 0.00)& (\textbf{0.04}, 0.00)\\ \cline{1-7}
 					Device 2	 &    (\textbf{0.05}, 0.00)	& (\textbf{0.06}, 0.00) & (\textbf{0.00, 1.00})& (\textbf{0.06}, 0.00) & (\textbf{0.13}, 0.00) & (\textbf{0.08}, 0.00)\\ \cline{1-7}
 					Device 3	&  	(\textbf{0.03}, 0.00)	& (\textbf{0.03},0.00) & (\textbf{0.06}, 0.00) & (\textbf{0.00, 1.00}) & (\textbf{0.06}, 0.00) & (\textbf{0.04},0.00)\\ \cline{1-7}
 					Device 4	 & (\textbf{0.04}, 0.00)& (\textbf{0.05},0.00) & (\textbf{0.13}, 0.00) & (\textbf{0.06}, 0.00) & (\textbf{0.00, 1.00}) & (\textbf{0.08}, 0.00)\\ \cline{1-7}
 					Device 5	 &   (\textbf{0.05}, 0.00) & (\textbf{0.04}, 0.00) & (\textbf{0.08}, 0.00) & (\textbf{0.04}, 0.00) & (\textbf{0.08}, 0.00) & (\textbf{0.00, 1.00})\\ \cline{1-7}
\end{tabular}
\end{table*}

%

\begin{table*}
\centering
\caption{Statistic and P-value for CAE model for all RF devices.}
\label{table:CAE1024_data_KS_Statistic}
\begin{tabular}{|c|c|c|c|c|c|c|}
\hline
	 		& \multicolumn{6}{c|}{RF Device} 	 \\ \cline{2-7}

{Device of Interest}		&  Device 0	& 	Device 1	&  Device 2	&  Device 3	&  Device 4	&  Device 5  \\\hline
 Device 0 		&    (\textbf{0.000, 1.000}) 	& (\textbf{0.313}, 0.000) & (\textbf{0.636}, 0.000) & (\textbf{0.555}, 0.000) & (\textbf{0.771}, 0.000)& (\textbf{0.627}, 0.000)		\\ \cline{1-7}
                			Device 1	&  	 (\textbf{0.312}, 0.000) 	& (\textbf{0.000, 1.000})& (\textbf{0.710}, 0.000) & (\textbf{0.334}, 0.000) & (\textbf{0.651}, 0.000)& (\textbf{0.380}, 0.000)\\ \cline{1-7}
 					Device 2	 &    (\textbf{0.635}, 0.000)	& (\textbf{0.685}, 0.000) & (\textbf{0.000, 1.000})& (\textbf{0.801}, 0.000) & (\textbf{0.198}, 0.000) & (\textbf{0.801}, 0.000)\\ \cline{1-7}
 					Device 3	&  	(\textbf{0.552}, 0.000)	& (\textbf{0.331},0.000) & (\textbf{0.801}, 0.000) & (\textbf{0.000, 1.000}) & (\textbf{0.979}, 0.000) & (\textbf{0.379},0.000)\\ \cline{1-7}
 					Device 4	 & (\textbf{0.778}, 0.000)& (\textbf{0.653},0.000) & (\textbf{0.236}, 0.000) & (\textbf{0.972}, 0.000) & (\textbf{0.000, 1.000}) & (\textbf{1.000}, 0.000)\\ \cline{1-7}
 					Device 5	 &   (\textbf{0.647}, 0.000) & (\textbf{0.381}, 0.00) & (\textbf{0.801}, 0.000) & (\textbf{0.384}, 0.000) & (\textbf{1.000}, 0.000) & (\textbf{0.000, 1.000})\\ \cline{1-7}
\end{tabular}
\end{table*}

%

\begin{table*}
\centering
\caption{Statistic and P-value for convolutional autoencoder model for all RF devices at different SNR levels (Device of interest: 5).}
\label{table:NoiseCAE1024_data_KS_Statistic}
\begin{tabular}{|c|c|c|c|c|c|c|}
\hline
	 		& \multicolumn{6}{c|}{RF Device} 	 \\ \cline{2-7}

{Noise level}		&  Device 0	& 	Device 1	&  Device 2	&  Device 3	&  Device 4	&  Device 5  \\\hline
 0 dB 		&    (\textbf{1.000}, 0.000) 	& (\textbf{0.999}, 0.000) & (\textbf{1.000}, 0.000) & (\textbf{1.000}, 0.000) & (\textbf{1.000}, 0.000)& (\textbf{0.000, 1.000})		\\ \cline{1-7}
                			-1 dB	&  	 (\textbf{1.000}, 0.000) 	& (\textbf{0.857}, 0.000) & (\textbf{1.000}, 0.000) & (\textbf{1.000}, 0.00) & (\textbf{1.000}, 0.000)& (\textbf{0.000, 1.000})	\\ \cline{1-7}
 					-5 dB	 &    (\textbf{0.424}, 0.000)	& (\textbf{0.274}, 0.000) & (\textbf{1.000}, 0.000)& (\textbf{0.589}, 0.000) & (\textbf{1.000}, 0.00) & (\textbf{0.000, 1.000})	\\ \cline{1-7}
 					-10 dB	&  	(\textbf{0.452}, 0.000)	& (\textbf{0.879},0.000) & (\textbf{1.000}, 0.000) & (\textbf{0.497}, 0.000) & (\textbf{0.993}, 0.00) & (\textbf{0.000, 1.00})	\\ \cline{1-7}
 					-15 dB	 & (\textbf{0.886}, 0.000)& (\textbf{0.932},0.000) & (\textbf{0.214}, 0.000) & (\textbf{0.657}, 0.000) & (\textbf{0.999}, 0.000) & (\textbf{0.000, 1.000})	\\ \cline{1-7}
 					[0,-1,-5,-10,-15]dB	 &   (\textbf{0.652}, 0.000) & (\textbf{0.395}, 0.00) & (\textbf{0.820}, 0.000) & (\textbf{0.445}, 0.00) & (\textbf{0.999}, 0.000) & (\textbf{0.000, 1.000})	\\ \cline{1-7}
\end{tabular}
\end{table*}

\begin{table*}
\centering
\caption{Statistic and P-value for convolutional autoencoder model for all USRP RF devices at different SNR levels (Device of interest: 2).}
\label{table:NoiseCAE1024_USRPdata_KS_Statistic}
\begin{tabular}{|c|c|c|c|c|c|}
\hline
	 		& \multicolumn{5}{c|}{RF Device} 	 \\ \cline{2-6}

{Noise level}		&  Device 0 & 	Device 1	&  Device 2	&  Device 3	&  Device 4 \\\hline
 -10 dB 		&    (\textbf{0.652}, 0.000) 	& (\textbf{0.557}, 0.000) & (\textbf{0.000}, 1.000) & (\textbf{0.580}, 0.000) & (\textbf{0.515}, 0.000)\\ \cline{1-6}
 
                			-8 dB &  (\textbf{0.607}, 0.000) 	& (\textbf{0.417}, 0.000) & (\textbf{0.000}, 1.000) & (\textbf{0.475}, 0.000) & (\textbf{0.435}, 0.000)\\ \cline{1-6}
                			
 					-6 dB & (\textbf{0.447}, 0.000)	& (\textbf{0.302}, 0.000) & (\textbf{0.000}, 1.000)	& (\textbf{0.290}, 0.000) & (\textbf{0.295}, 0.000)\\ \cline{1-6}
 					
 					-4 dB &  (\textbf{0.508}, 0.000)	& (\textbf{0.340}, 0.000) & (\textbf{0.000}, 1.000) & (\textbf{0.337}, 0.000) & (\textbf{0.243}, 0.00) \\ \cline{1-6}
 					
 					-2 dB & (\textbf{0.800}, 0.000)	& (\textbf{0.700}, 0.000) & (\textbf{0.000}, 1.000) & (\textbf{0.680}, 0.000) & (\textbf{0.701}, 0.000)\\ \cline{1-6}
 					
 					0dB	 &  (\textbf{0.587}, 0.000)	& (\textbf{0.366}, 0.000) & (\textbf{0.000}, 1.000) & (\textbf{0.408}, 0.000) & (\textbf{0.209}, 0.000)\\ \cline{1-6}
 					
 					2 dB	 & (\textbf{0.680}, 0.000)	& (\textbf{0.605}, 0.000) & (\textbf{0.000}, 1.000)	& (\textbf{0.650}, 0.000) & (\textbf{0.424}, 0.000)\\ \cline{1-6}
 					
 					4 dB	&  (\textbf{0.615}, 0.000)	& (\textbf{0.534}, 0.000) & (\textbf{0.000}, 1.000) & (\textbf{0.512}, 0.000) & (\textbf{0.482}, 0.000)\\ \cline{1-6}
 					
 					6 dB	 & (\textbf{0.525}, 0.000)	& (\textbf{0.535}, 0.000) & (\textbf{0.000}, 1.000) & (\textbf{0.520}, 0.000) & (\textbf{0.525}, 0.000) \\ \cline{1-6}
 					
 					8 dB	 & (\textbf{0.487}, 0.000) 	& (\textbf{0.512}, 0.000) & (\textbf{0.000}, 1.000) & (\textbf{0.489}, 0.000) & (\textbf{0.457}, 0.000)\\ \cline{1-6}
 					
 					10dB & (\textbf{0.373}, 0.000) 	& (\textbf{0.499}, 0.000) & (\textbf{0.000}, 1.000) & (\textbf{0.467}, 0.000) & (\textbf{0.462}, 0.000) \\ \cline{1-6}
 					
 	[-10,10]dB		& (\textbf{0.319}, 0.000) & (\textbf{0.356}, 0.000) & (\textbf{0.000}, 1.000) & (\textbf{0.339}, 0.000) & (\textbf{0.411}, 0.000)\\ \cline{1-6}
\hline
\end{tabular}
\end{table*}

In summary, it is evident that DAC approach is successful at exploiting device inherent features. In an authentication scenario, if the K-S statistic and the \textit{p}-value between the DAC of the data received by a sending device, and that of the reconstruction error produced by receiving device are 0 and 1 respectively, then that device is an authorized device. Otherwise that device can be flagged as an intruder. Furthermore, it has been shown that the model is robust to varying SNR levels in the transmitted signal. 

\section{Discussion}
\label{sec:discussion}
According to Baldini in~\cite{paper6_Baldini2017} the major requirements for fingerprinting as adopted from the biometric domain are:
\begin{enumerate}
\item \textbf{universality}, meaning that every device must be identifiable by the characteristics of its built-in electrical components.
\item \textbf{uniqueness}, meaning that no two devices should have the same fingerprint signatures or physical characteristics.
\item \textbf{permanence}, which requires that the features must be invariant to time and environmental conditions.
\item \textbf{collectability}, which requires that the characteristics must me quantitatively measurable
\end{enumerate}

All requirements may not be satisfied simultaneously or to the same extent for all components in an IoT device. Furthermore, in current state-of-the-art identification approaches, the features may be time varying, dependent on the environment, or in some cases may not be adequately discriminatory for the identification of the device~\cite{paper6_Baldini2017}. Some methods adopt a statistical approach to feature extraction, others methods formulate RF fingerprinting as a classification problem and apply supervised machine and deep learning techniques on the statistical features. There are methods that propose automatic feature extraction from the raw RF signals using deep learning and others adopt unsupervised learning methods.


In~\cite{paper36_Lukacs2015} the authors apply a statistical approach called ``Radio Frequency Distinct Native Attribute (RF-DNA)" for passive interrogation of microwave devices. The RF-DNA is a novel method developed by the Air Force institute of Technology in 2006. RF-DNA extracts the variance, skewness and kurtosis features from three statistical properties of the signal namely: the instantaneous amplitude, phase and frequency of the received signal. The statistical features are then concatenated to form the fingerprint or ``DNA''. Multiple discriminant analysis (MDA) and a maximum likelihood classifier (MLC) were applied for dimension reduction and classification respectively. The authors also applied non-parametric random forest and Adaboost classifiers using RF-DNaA features in~\cite{paper20_patel2015} and obtained better results than the MLC. Extending this work, Bihl et. al~\cite{paper14_bihl2016} applied multiple-discriminant analysis to reduce the dimensions of the RF-DNA and showed that better results could be obtained. The authors in~\cite{paper33_wang2017} adopt the same RF-DNA method to address physical layer security for cognitive radio networks.

The authors in~\cite{paper46_ali2019} employed a symbol-based statistical RF fingerprinting approach for identification of fake base station (FBS) in a cellular network. In their work, leverage is made on the fact that the amplitude and phase errors introduced in the transmitted signal will be larger for a FBS compared to a real base station (RBS). The non-linearity introduced by the power amplifier  of several software defined radio in the FBS is measured at the user equipment  using a second-order symbol-based error vector magnitude (EVM) approach. After this, the structure of the noise in the signal the UE receives is determined using the kurtosis; specifically the second and fourth order moments. The authors assert that the kurtosis on the magnitude of the noise structure is an effective indicator that can be used to identify a FBS.

%
%
Patel in~\cite{paper27_Patel2015} proposed the use of non-parametric methods for feature generation such as mean, median, mode and trend rather parametric methods such as variance, standard deviation, skewness and kurtosis. By testing this using a random forest classifier, he asserts that features obtained from non-parametric features result in improved classification. In~\cite{paper48_ureten2007} the authors tackle the problem of identifying a node by its fingerprint. In their work they use features extracted from the complex amplitude and phase angle of the WiFi signals. Hilbert transform is applied as data prepossessing. They however discard the phase profiles and use the amplitude profiles for feature extraction. Principal component analysis (PCA) was used to reduce the dimension and the reduced data is fed into a neural network for classification.

%
%
There are many other works that follow the approach of statistically generating RF fingerprints and feeding them into a classification model such as~\cite{paper26_nouichi2019}~\cite{paper18_baldini2017}~\cite{paper17_Chen2017}. Similar to this work some work such as~\cite{paper24_jafari2018} actually perform classification on the raw RF data. These works do not consider that training data of some classes are unavailable. Moreover, in most of the works, the proposed approaches are tested on data obtained at singular noise levels, which does not always depict real life scenario.

An anomaly detection technique based on a deep predictive coding neural network, for analyzing RF spectrum in wireless systems was proposed in~\cite{Tandiya2018}. In their work, frequency-domain data obtained from time-domain data were stored  as sequential 2D images. The image sequences were then fed into a deep learning video predictor which attempts to predict the next frame from previous frames. Anomaly detection is triggered when there is a deviation between the actual and predicted spectrum behavior. In~\cite{Sixing2009}, an anomaly identification method in temporal-spectral data was proposed. They generate models from historical data and compare the historical data with real-time data for intrusion detection. These approaches do not require RF data from all classes. However, there is a challenge of specifying what normal system behavior means, as well as defining an appropriate threshold. 

There have been recent works that have applied generative models. The very recent work by Roy et.al in~\cite{paper59_roy2019} proposes the use of GANs for the detection of rogue RF transmitters. They use the generator model of the GAN to learn the sample space of the IQ values of the authorized transmitters. They then generate fake signals that mimic the transmissions of the authorized transmitters from the learned representation. The authors test their model with fake data generated by the model, as opposed to actual data from an unauthorized device. Furthermore, data was collected at one SNR level of 45dB which infers very strong signal. Also, more recently the authors in~\cite{Rajendran2018} introduced a method called spectrum anomaly detector with interpretable features (SAIFE) which is an anomaly detection framework for wireless spectrum based on adversarial autoencoders. In their work, the authors make use of power spectral density (PSD) data. They showed that their model can achieve a constant false alarm rate in an unsupervised setting, whereas in a semi supervised setting, the model is capable of learning intuitive features such as class center frequency and signal bandwidth. Their approach achieves very impressive results while being exposed to just 20$\%$ of labeled data samples.

To the best of our knowledge, most of the existing methods in the literature do not consider the unavailability of data from some classes (e.g., intruder) during training. For the more recent works that use generative models and consider identifying novel classes, the models are trained and tested with data of only one SNR level which may not be representative of certain real life scenarios. Furthermore, to the best of our knowledge, this is the first work that combines deep learning with information theoretic approach to authentication in the RF domain. Our approach has been verified on data which contains different SNR levels to represent the real life effects of varying channel conditions, device mobility and noise.

\section{Conclusions}
\label{sec:conclusions}
In this work, we propose a novel framework for intrusion detection based on RF fingerprinting using deep learning. Specifically, the problem of identifying an authorized device or an intruder  from a set of devices of the same make, model and manufacturer sending the exact same information is considered, and a novel concept of Device Authentication Code (DAC) is proposed. In the proposed framework, an autoencoder is used to automatically extract features from the RF traces, and the  reconstruction error is used as the DAC, and this DAC is unique to each device and the particular message of interest. Then Kolmogorov-Smirnov (K-S) test is used to match the distribution of the reconstruction error generated by the autoencoder and the received message, and the result will determine whether the device of interest belongs to an authorized user. We validate this concept on two experimentally collected RF traces from six ZigBee and five universal software defined radio peripheral (USRP) devices, respectively. Experimental results demonstrate that DAC is able to prevent device impersonation by extracting salient features that are unique to any wireless device of interest and can be used to identify RF devices. 
Furthermore, we show that our method is robust to changes in channel conditions, mobility and varying signal strength. 
It is worth noting that the proposed method does not need the RF traces of the intruder during model training yet be able to identify devices not seen during training, which makes it practical.

\section{Acknowledgment}
\label{sec:acknowledgement}
This research work is supported in part by the U.S. Dept. of Navy under agreement number N00014-17-1-3062 and the U.S. Office of the Under Secretary of Defense for Research and Engineering (OUSD(R\&E)) under agreement number FA8750-15-2-0119. The U.S. Government is authorized to reproduce and distribute reprints for governmental purposes notwithstanding any copyright notation thereon. The views and conclusions contained herein are those of the authors and should not be interpreted as necessarily representing the official policies or endorsements, either expressed or implied, of the Dept. of Navy  or the Office of the Under Secretary of Defense for Research and Engineering (OUSD(R\&E)) or the U.S. Government.

\bibliographystyle{IEEEtran}
\bibliography{IOTSecurityJeff}
\end{document}